\documentclass[twocolumn,english,showpacs,aps,prl]{revtex4}
\usepackage[T1]{fontenc}
\usepackage[latin1]{inputenc}
\usepackage{float}
\usepackage{amsmath}
\usepackage{graphicx}
\usepackage{amssymb}

\makeatletter

\usepackage{graphicx}
\usepackage{amsfonts}

\usepackage{babel}
\makeatother

\begin{document}

\title{Critical slowing down and fading away of the piston effect in porous media}
\author{Bernard Zappoli *}
\affiliation{{*} CNRS-PMMH, ESPCI, 10 rue Vauquelin 75005 Paris, France}

\author{Rapha\"{e}l Cherrier** and Didier Lasseux**}
\affiliation{{**} CNRS-TREFLE, ENSAM, Esplanade des Arts et M\'{e}tiers, 33405 Talence France}

\author{Jalil Ouazzani*** and Yves Garrabos***}
\affiliation{{***}ESEME - ICMCB - CNRS - Université Bordeaux I - 87, avenue du Docteur Schweitzer, F 33608 Pessac Cedex France.}

\date{\today}

\begin{abstract}
We investigate the critical speeding up of heat equilibration by the piston effect (PE) in a nearly supercritical van der Waals (vdW) fluid confined in a homogeneous porous medium.
We perform an asymptotic analysis of the averaged linearized mass, momentum and energy equations to describe the response of the medium to a boundary heat flux. While nearing the critical point (CP), we find two universal crossovers depending on porosity, intrinsic permeability and viscosity. Closer to the CP than the first crossover, a pressure gradient appears in the bulk due to viscous effects, the PE characteristic time scale stops decreasing and tends to a constant. In infinitly long samples the temperature penetration depth is larger than the diffusion one indicating that the PE in porous media is not a finite size effect as it is in pure fluids. Closer to the CP, a second cross over appears which is characterized by a pressure gradient in the thermal boundary layer (BL). Beyond this second crossover, the PE time remains constant, the expansion of the fluid in the BL drops down and the PE ultimately fades away.
\end{abstract}


\maketitle

Near critical pure fluids are Newtonian, viscous and heat conducting fluids.
Their diverging compressibility allows a fourth energy transport mechanism, called the piston effect (PE), to become prominent \cite{1, 2, 3}.
The fluid contained in thermal boundary layers (BL) formed along heated walls expands, generates acoustic compressions waves which flash back and forth in the finite size sample fluid, and induces an adiabatic heating  of the bulk much faster than it would be by heat diffusion.
The characteristic time of heat equilibration tends to zero to 0 as $\tau^{3 / 2}$ for a vdW fluid, where $\tau$ is the dimensionless distance to the critical point (CP) defined as $\left( T - T_C \right) / T_C$, $T$ being the temperature of the fluid and $T_C$ its critical temperature. 
This remarkable finite size mechanism contrasts with the critical slowing down of heat diffusion.
The homogeneous character of both temperature and pressure (except in the thermal BL for temperature) are typical features of the PE.
Viscous effects are always negligible in bulk supercritical fluid although accounting for a strong divergence of the bulk viscosity \cite{4} may lead to small scale thermal boundary layer gradients.

In this letter, we address the 1D problem of heat transport in a near critical fluid contained in a porous medium .
We consider the gravity vector oriented perpendicularly to the planar boundaries of the thin sample (1-D approximation) or that gravity is equibrated by inertia forces as in space laboratories.
We basically find that pressure gradients which build up progressively while nearing the CP (first in the bulk and then in the thermal BL) considerably weaken and slow down the PE which still exists in infinite media but ultimately vanishes.
The configuration under study is that of two parallel and infinite planes distant by $L$ limiting a homogeneous rigid porous medium saturated by a near critical fluid initially at rest, at thermodynamic equilibrium, at its critical density $\rho_c$ and set just above its critical temperature so that $\tau \langle\langle 1$.
This defines the reference conditions.
At $t^{\prime} \geqslant 0$, the left boundary is heated with a flux $\phi(t^{\prime})$ while the right one remains thermally insulated. In the rest of this paper, subscripts $c$ and $0$ are used to denote quantities in the fluid phase at the CP and in the reference conditions respectively.

As supercritical fluids can be described as compressible, viscous, heat conducting and newtonian van der Waals fluids, we consider the averaged transport equations over an elementary volume containing a large number of pores to be identical to that obtained for an ideal gas in a porous medium.
The masss conservation equation is given in \cite{5}.
The momentum balance is expressed by the unsteady form of the Darcy equation \cite{6} which needs to be valid that the pressure gradient at the pore scale is small compared to that at the macroscopic scale.
The energy balance which acounts for the work of pressure forces is similar to that of pure fluids, since  the one temperature model as derived in \cite{7} leads to a solid contribution which only modifies the heat capacity and the thermal conductivity of the pure fluid.

At time $t'=0$, heat deposited at $x^{\prime} = 0$ diffuses into the medium and we derive the asymptotic analysis of the coupled equations.
To this end, we divide the domain into two sub systems: a thin diffusion BL of thichness $\delta $, $\delta \langle\langle 1$ when referred to the sample length, and the bulk.
The characteristic heat flux at the wall is such that the averaged density, velocity, pressure and temperature of respective first orders of magnitude $\tilde{\eta}_\rho , \tilde{\eta}_u , \tilde{\eta}_P$ and $\tilde{\eta}_T$ in the BL undergo small perturbations with respect to the reference conditions.
If we denote $\theta = \varsigma t$, $ \varsigma\langle\langle 1 $, $\theta$ is referred to a time scale which is  $\varsigma^{-1}$ times longer than the sample acoustic characteristic time which defines the non dimensional time $t$ as indicated below.
The non-dimensional, rescaled, governing equations in the BL are:
\begin{eqnarray}
	& \varsigma \tilde {\eta }_\rho \,\,\tilde {\rho}_\theta + \delta^{-1} \tilde{\eta}_u \tilde{u}_z =0 &
	\label{eq1}
	\\
	& \varsigma\tilde \eta_u \tilde u_\theta + \delta^{-1} \gamma_0^{-1} \tilde{\eta}_P \tilde{P}_z = -\left( 4 / 3 \right)
	\left( \varepsilon / K \right) \tilde{\eta}_u \tilde{u} &
	\label{eq2}
\end{eqnarray}
\begin{equation}
	\begin{array}{rl}
		\left( {\rho C_v} \right)^\ast \varsigma \tilde{\eta}_T \tilde{T}_\theta = & - \frac{3}{2} \delta^{-1} \tilde{\eta}_u \tilde{u}_z 
		\\
		& + \left( \frac{\gamma_0}{Pr_0 \left(\gamma_0 - 1\right)} \right) \varepsilon \delta^{-2} \tilde{\eta}_T \tau ^{-0.5} \tilde{T}_{zz}
	\end{array}
	\label{eq3}
\end{equation}
\begin{eqnarray}
	&\tilde{\eta}_P \tilde{P} = \left( \frac{3}{2} \right) \tilde{\eta}_T \tilde{T} + \frac{9}{4} \tau \tilde{\eta}_\rho \tilde{\rho} &
	\label{eq4}
	\\
	& - \delta^{-1} \tau^{-0.5} \tilde{\eta}_T \tilde{T}_z = \varphi W (\theta) & \text{ at } z = 0
	\label{eq5}
	\\
	& \tilde{\rho} = \tilde{u} = \tilde{T} = \tilde{P} = 0 & \text{ at } \theta = 0 
	\label{eq6}
	\\
	& \tilde{u} \left( z,\theta \right) = 0 & \text{ at } z = 0
	\label{eq7}
\end{eqnarray}

In these equations, dimensionless temperature, density, pressure and velocity are defined from the corresponding dimensional quantities denoted with a prime as follows:
\begin{itemize}
	\item $T \left( x,t \right) = \frac{T'(x',t')}{T_C } = 1 + \tau + \tilde{T} \left( z,\theta \right) \tilde{\eta}_T$;
	\item $\rho \left( x,t \right) = \frac{\rho'(x',t')}{\rho _C} = 1 + \tilde{\rho} \left( z,\theta \right) \tilde{\eta}_\rho$;
	\item $P\left( x,t \right) = \frac{P'(x',t')}{\gamma_0 c_0^2} = P_0 + \tilde{\eta}_P \tilde{P} \left( z,\theta \right)$
\end{itemize}
where $r$, $\gamma_0$, $c_0$ are the ideal gas constant per unit mass, the ratio of specific heats and the sound velocity in the corresponding ideal gas; 
$u\left( x,t \right) = \frac{u'(x',t')}{c_0} = \tilde{\eta}_u \tilde{u} \left( z,\theta \right)$.

The dimensionless boundary layer coordinate is given by: 
$z = \left( \frac{x^{\prime}}{L} \right) \delta^{-1} = x \delta^{-1}$
 where $x$ is the dimensionless space variable at the sample length scale while the time variable $t$ is referred to the acoustic time and is defined by $t = \frac{c_0 t^{\prime}}{L}$.
$\varepsilon $ is the ratio, far from the critical temperature, of the sample acoustic time $\frac{L}{c_0}$ to the sample viscous diffusion time $\frac{L^2}{\nu_0 }$, $\nu_0 $ being the kinematic viscosity of the fluid in the reference conditions.

For usual critical fluids $\varepsilon$ is a small parameter of order $10^{-8}$; larger values correspond to more viscous supercritical fluids or to thinner samples.
The Darcy number $K$ is the intrinsic permeability $K^\prime$ divided by the squared characteristic length of the sample and by the porosity $\varepsilon_p$ : 
$K = {K^{\prime}} \mathord{\left/ {\vphantom {{K'} {\left( {\varepsilon _p \,L^2} \right)}}} \right. \kern-\nulldelimiterspace} {\left( {\varepsilon _p \,L^2} \right)}$. $K'$ is proportional to the squared characrteristic pore size and the porosity is the ratio of the total pore volume to the total volume of the sample.

The ratio $\varepsilon / K$ can vary over several orders of magnitude (typically from $10^{-3}$ to $10^4)$ while $\varepsilon^2 / K$ is always much smaller than unity.
Equation (\ref{eq3}) is the internal energy balance equation which shows the work of pressure forces $(-\left( 3 / 2 \right) \delta^{-1} \tilde{\eta}_u \tilde{u}_z)$ and heat diffusion $(\varepsilon \left( \frac{\gamma_0}{Pr_0 \left( \gamma_0 - 1 \right)} \right) \delta^{-2} \tilde{\eta}_T \tau^{-.05} \tilde{T}_{zz})$ terms.
The near critical fluid thermal conductivity, $\Lambda^F$, is described by the mean field theory \cite{8} i.e. 
$\Lambda^F = \frac{\Lambda^{\prime F}}{\Lambda_0^{\prime F}} = \frac{\Lambda_b^{\prime}}{\Lambda_0^{\prime F}} + \tau^{-0.5} \cong \tau^{-0.5}$ where $\Lambda_b^{\prime}$ is its background value and $\Lambda_0^{\prime F}$ its critical amplitude.

In (\ref{eq3}), we have neglected the solid heat conductivity compared to the diverging one of the near-critical fluid and the Maxwell approximation \cite{9} was used to derive the equivalent heat conductivity of the porous medium to give $\left( \frac{2 \varepsilon_p}{3 - \varepsilon_p} \right) \Lambda^F = \left( \frac{2 \varepsilon_p}{3 - \varepsilon_p} \right) \tau^{-0.5}$.
$\Pr_0$ is an equivalent Prandtl number defined by $\Pr_0 = \frac{\varepsilon_p \nu_0}{D_{T0}}$ where $D_{T0}$ is a reference thermal diffusivity defined by $D_{T0} = \frac{2 \varepsilon_p}{3 - \varepsilon_p} \frac{\Lambda_0^{\prime F}}{\rho_c c_{p0}^{\prime}}$ where $c_{p0}^{\prime} = \frac{\gamma_0}{\gamma_0 - 1} r$ is the specific heat at constant pressure of the ideal gas.

One should note that the mean field theory yields a $\tau^{-1}$ divergence of the heat capacity at constant pressure and thus a heat diffusivity which goes to zero as $\tau^{0.5}$.
In the left hand side of (\ref{eq3}), the term $\left( \rho C_v \right)^\ast$ is the equivalent non-dimensional heat capacity at constant volume given by $\left( \rho C_v \right)^\ast = \frac{1}{\gamma_0 - 1} + \frac{1 - \varepsilon_p}{\varepsilon_p} \rho_s c_s$, where 
$c_{v 0} = \frac{c_{v0}^{\prime}}{r} = \frac{1}{\gamma_0 - 1}$ is the dimensionless heat capacity at constant volume of the near-critical fluid 
considered as constant because of its weak divergence as $\tau^{-0.11}$ when nearing the CP and $\rho_s c_s$ the dimensionless heat capacity of the solid defined by $\rho_s c_s = \frac{\rho_s^{\prime} c_s^{\prime}}{\rho_c r}$.

Equation (\ref{eq4}) is the linearized vdW equation of state.
Equation (\ref{eq5}) is the boundary condition at the left wall ($z = 0$) where $0 \leqslant W (\theta) \leqslant 1$ is the time modulation of the flux deposited at this boundary.
The dimensionless parameter $\varphi$ in boundary condition (5) is the characteristic heat flux defined as $\varphi = \frac{W_{\max} L}{T_c^{\prime} \Lambda_0^{\prime F}}$ where $W_{\max}$ is the maximum value over time of the deposited flux at the boundary.
Equation (\ref{eq6})is the initial conditions.

In the bulk region, the governing equations for the perturbed variables $\bar{\rho} \left( x,\theta \right)$, $\bar{u} \left( x,\theta \right)$, $\bar{P} \left( x,\theta \right)$ and $\bar {T} \left( x,\theta \right)$ are similar to (\ref{eq1}-\ref{eq4}) except the space variable is $x$, 
$\delta $ is replaced by 1, the orders of magnitude are respectively, $\bar{\eta}_\rho , \bar{\eta}_u , \bar{\eta}_P , \bar{\eta}_T$ and the boundary and initial conditions are: 
\begin{eqnarray}
	& \bar{T}_x = 0 & \text{ at } x = 1
	\label{eq8}
	\\
	& \bar{\rho} = \bar{u} = \bar{T} = \bar{P} = 0 & \text{ at } \theta = 0
	\label{eq9}
	\\
	& \bar{u} \left( x,\theta \right) = 0 & \text{ at } x = 1
	\label{eq10}
\end{eqnarray}

Matching conditions between the bulk and the BL regions close the equations, namely 
\begin{equation}
	\mathop{\lim} \limits_{z \to \infty} \tilde{\eta}_F \tilde{F} \left( z,\theta \right) = \mathop{\lim} \limits_{x \to 0} \bar{\eta}_F \bar{F} \left( x,\theta \right) \, ,
	\label{eq11}
\end{equation}
where $F$ is a physical unknown.

Finding an approximate solution of the equations first needs determining the orders of magnitude functions.
Boundary condition \cite{5} for temperature implies that $\tilde{\eta}_T
\delta \tau^{0.5} \varphi = \tilde{\eta}_T$.
The coupling of the thermodynamic variables through the equation of state \cite{4}) gives $\tilde{\eta}_T\ = \tilde{\eta}_P = \tau \tilde{\eta}_\rho$ while the mass conservation equation gives $\varsigma \tilde {\eta }_\rho = \delta^{-1} \tilde{\eta}_u$ .
Moreover, as the matching between diffusion and the work of pressure forces in the boundary layer (diffusion matches with the bigest term in the equation) \cite{10} is expressed by $\tilde{\eta}_u \delta^{-1} = \varepsilon \delta^{-2} \tau^{-0.5} \tilde{\eta}_T$, the previous equations finally link space and time scales through the relation $\varsigma \delta^2 = \varepsilon \tau^{0.5}$
and give the order of magnitude $\tilde{\eta}_u = \varepsilon \varphi$ of the velocity in the BL 
The above orders of magnitude allow writing the Darcy equation as
\begin{eqnarray}
	& \varsigma \tilde{u}_\theta + \varepsilon^{-1} \tau^{0.5} \tilde{P}_z / \gamma _0
	= - 4 / 3 \left( \varepsilon / K \right) \bar {u} &
	\label{eq12}
	\\
	& K  / \varepsilon \langle\langle 1 &
	\label{eq13}
\end{eqnarray}

If condition (\ref{eq13}) is fulfilled, which will be supposed valid throughout the paper, and if $\tau \rangle\rangle \tau_{c2} = \left( \varepsilon^2 / K \right)^2$, which will be further commented later in the text, then the following hierarchy holds 
\begin{equation}
	\varsigma \langle\langle 1 \langle\langle \varepsilon / K \langle\langle \varepsilon^{-1} \tau^{0.5} \, 
	\label{eq15}
\end{equation}
and the Darcy equation in the BL reduces to $\tilde{P}_z = 0$.

The BL solution shows that all the variables go to a constant at the outer edge of the BL.
Considering the matching condition (\ref{eq11}) for the velocity, 
$\mathop{\lim}\limits_{z \to \infty} \tilde{\eta}_u \tilde{u} \left( z , \theta \right) = \mathop{\lim}\limits_{x \to 0} \bar{\eta}_u \bar{u} \left( x , \theta \right)$, and looking for a temperature homogeneization time scale (both temperature in the bulk and in the BL are of the same order of magnitude) thus gives: 
\begin{equation}
	\tilde{\eta}_T = \bar{\eta}_T = \delta \varphi \tau^{0.5} \text { and } \tilde{\eta}_u = \bar{\eta}_u = \varepsilon \varphi, .
	\label{eq16}
\end{equation}

By carrying the latter into the bulk energy balance and matching the transient term with the term representing the work of pressure forces (diffusion in the bulk being negligible), we obtain the BL thickness $\delta = \tau$ and thus the value of the time scale $\varsigma = \varepsilon \tau^{-3 / 2}$.

These orders of magnitude allow writing the Darcy's Equation in the bulk as: 
\begin{equation}
	 \frac{\varepsilon^2}{\tau^{3 / 2}} \, \bar{u}_\theta + \tau^{3 / 2} \, \frac{\bar{P}_x}{\gamma_0} = - \frac{4}{3} \left( \varepsilon^2 / K \right) \, \bar{u} 
	\label{eq17}
\end{equation}

If condition $\tau \rangle\rangle \tau_{c1} = \left( \varepsilon^2 / K \right)^{\frac{2}{3}}$ is verified, the following hierarchy holds$\frac{\varepsilon}{\tau^{3 / 2}} \langle\langle \frac{K}{\varepsilon} \langle\langle \frac{\varepsilon}{K} \langle\langle \frac{\tau^{3 / 2}}{\varepsilon}$ and Darcy's equation in the bulk phase also reduces to $\bar{P}_x = 0$.
Under these conditions, the governing equations are exactly the same as those for the pure supercritical fluid \cite{10,11} and the solution in the bulk displays homogeneous temperature, pressure and density.
The PE time scale $\varsigma^{-1}$, i.e. the time scale on which temperature in the sample is homogenized by thermo-acoustic heating goes to zero as $\tau ^{3 \mathord{\left/ {\vphantom {3 }} \right. \kern-\nulldelimiterspace} 2}$.
It is thus smaller than the heat diffusion time which goes to infinity as $\tau ^{-0.5}$.
This critical speeding up of heat equilibration contrasts with the critical heat diffusion slowing down.
In other words as far as condition $\tau \rangle\rangle \tau_{c1} = \left( \varepsilon^2 / K \right)^{\frac{2}{3}}$ is fulfilled, the PE is not affected by the porous matrix.
When nearing the CP, $\tau $ becomes first of the order of $\tau _{c1} $ and a pressure gradient builds up in the bulk phase.
Therefore, a first crossover to porous effects arises when the pressure gradient and the average viscous stress in  (\ref{eq17}) become of the same order of magnitude, i.e. when
\begin{equation}
	\tau \approx \tau_{c1} = \left( \frac{\varepsilon^2}{K} \right)^{\frac{2}{3}}
	\label{eq18}
\end{equation}
When written in dimensional parameters, this crossover value becomes 
\begin{equation}
	\frac{T^{\prime} - T_c}{T_c} = \frac{\left( \nu_0^\prime / c_0 \right)^{4 / 3}}{K^{\prime 2 /3}}
	\label{eq19}
\end{equation}
$\tau_{c1}$ it thus is a universal value, only depending on the fluid and solid matrix properties.

For carbon dioxide $\nu_0^{\prime} = 0.66 \, .10^{-7} \, m^2s^{-1}$, $c_0^{\prime} = 3 \, .10^2 \, m.s^{-1}$, and for reasonable values of the ratio
 ${K^{\prime}} \mathord{\left/ {\vphantom {{K'} {\varepsilon _P }}} \right. \kern-\nulldelimiterspace} {\varepsilon _P }=10^{-14}\,m^2$, the crossover value is at about $0.09 K$ away from the critical temperature. One should note that for more viscous critical fluids, this cross over value would be much farther from the critical temperature.

The solution of the system of equations when nearing the CP is obtained by performing a crossover type solution, e.g. by repeating the above analysis when $\tau$ tends to zero while $\tilde{\tau} = \tau / \left( \varepsilon^2 / K \right)^{2 / 3}$ is kept constant.
Doing so leads to the following orders of magnitude: 
\begin{equation}
	\begin{array}{lcr}
		& \tilde{\eta}_T = \tilde{\eta}_p = \bar{\eta}_T = \bar{\eta}_P = \bar{\eta}_\rho = \tau_{c1}^{3 / 2} \varphi &
		\\
		\text{and} & & ,
		\\
		& \tilde{\eta}_u = \bar{\eta}_u = \varepsilon \varphi &
	\end{array}
	\label{eq20}
\end{equation}
as well as to the following time scale parameter $\varsigma = K / \varepsilon$ and thermal boundary layer thickness $\delta = \left( \varepsilon^2 / K \right)^{\frac{2}{3}}$.

As beyond the crossover we have per definition 
\begin{eqnarray}
	& \tau^{3 / 2} \langle\langle \varepsilon^2 / K &
	\nonumber
	\\
	\text{which implies} & & . \label{eq21}
	\\
	& \theta = \frac{\varepsilon t}{\varepsilon^2 / K} \langle\langle \ \varepsilon t / \tau^{3 / 2} &
	\nonumber
\end{eqnarray}

As $\theta_{PE} = \varepsilon t / \tau^{3 / 2}$ is the time variable corresponding to the PE in the pure fluid, we can conclude that the PE time scale in porous media is longer than the one in the pure fluid. The effect of the solid matrix is thus to slow down the temperature homogenization whose charateristic time tends to a constant equal to $\varepsilon / K$ when referred to the acoustic time  instead of being shorter and shorter as in pure fluid. We call the phenomenon the porous saturation of the critical speeding up.

Combining (\ref{eq1}-\ref{eq5}) with the relevant orders of magnitude leads to a diffusion equation for density in the BL whose solution can be written as follows 
\begin{equation}
	\tilde{\rho} \left( \theta ,z \right) = - \frac{2}{3} A \int\limits_0^\theta {K_T \left( \theta - u , z \right) W(u) du}
	\label{eq25}
\end{equation}
where 
$A \left( \tilde{\tau} \right)
= \frac{4}{3} \left( \gamma_0 / {\gamma_0 - 1} \right) \Pr_0^{-1} \tilde {\tau }^{0.5}
\equiv A \tilde{\tau}^{0.5}$
is the diffusion coefficient and
\begin{equation}
	K_T \left( u \right) 
	= \frac{1}{\sqrt{\pi A \left( \tilde{\tau} \right)} u}  
	\exp \left( \frac{- z^2}{4 A \left( \tilde{\tau} \right)} u \right)
	\label{eq26}
\end{equation}
is a normalized Gaussian diffusion kernel.
Going now to the bulk equations we obtain a diffusion equation for temperature whose solution is: 

\begin{equation}
	\bar{T} \left( \theta ,x \right) = \frac{A}{\left( \rho C_v \right)^\ast} \int\limits_0^\theta \bar{K}_T \left( \theta -u ,x \right) W(u) du \, ,
	\label{eq27}
\end{equation}
where 
\begin{equation}
	\begin{array}{rl}
		\bar{K}_T \left( u, x \right) = 1 + 2 \sum\limits_{n = 1}^\infty 
		& \left( -1 \right)^n 
		\\ & \times \exp \left( -n^2 \pi^2 B(tilde{\tau}) u \right)
		\\ & \times \cos \left[ n \pi (1 - x) \right] 
	\end{array}
	\label{eq28}
\end{equation}
is the kernel associated with temperature and 
$B \left( \tilde{\tau} \right) = \left( 27 / 16 \gamma_0 \left( \rho C_v \right)^\ast \right) \tilde{\tau}^{3/2}$ 
is an effective diffusion coefficient.
The first correction brought by the presence of the porous medium appears in the $n=1$ term in the series (\ref{eq28}).

In the limit $\tilde{\tau} \langle\langle 1$, $\bar{K}_T$ becomes the diffusion kernel $\bar{K}_T \left( u \right)= \left( \frac{1}{\sqrt{\pi B \left( \tilde{\tau} \right)} u } \right)\exp \left( - \frac{z^2}{4 B \left( \tilde{\tau} \right)} u \right)$ which indicates that there is an effective energy penetration depth $\sqrt {B\left( {\tilde {\tau }} \right)} $ in an infinite porous medium saturated by a near critical fluid, larger than the heat diffusion penetration depth on the same time scale.
An effective energy transfer process in an infinite bulk thus exists whose penetration depth, $\sqrt {B\left( {\tilde {\tau }} \right)} $ is longer than the diffusion one on the same time scale. This energy transfer, faster than diffusion in a convection free infinite medium, can be interpreted as due to the partial reflection of the heating, boundary layer generated, acoustic waves on the porous matrix. 
The thermoacoustic origin of this finite size effect could be confirmed by looking at the acoustic propagation modes as was done for the PE in a pure supercritical fluid \cite{11}.

When nearing the CP much closer than $\tau_{c1}$, the already-mentioned value $\tau \approx \tau_{c2} = \left( \varepsilon^2 / K \right)^2$ is reached for which a pressure gradient appears in the BL (see (\ref{eq12})).
Although the small value of the distance to the CP of this second crossover for usual supercritical fluids ($\tau_{c2} \approx\ 10^{-5} K$ for carbon dioxyde for L=10 mm) may suggest severe limitations to the Navier Stokes approach, a new crossover analysis leads to a diffusion equation for density in the boundary layer whose solution,
\begin{equation}
	\begin{array}{rl}
		\tilde{\rho} \left( \theta ,z \right) =&  - \frac{2}{3} A_2 \left( \tilde{\tau}_2 \right) \tilde{\tau}_2^{-1 / 2}
		\\
		& \int\limits_0^\theta K_{T2} \left( \theta - u ,z \right) W \left( u \right) du \, ,
	\end{array}
	\label{eq29}
\end{equation}
where the associated diffusion coefficient  
\begin{equation}
	A_2 \left( \tilde{\tau}_2 \right) = \left[ \frac{16}{27} \gamma_0 \tilde{\tau}_2^{-3/2} +A \left( \tilde{\tau}_2 \right)^{-1} \right]^{-1}
	\label{eq30}
\end{equation}
and diffusion kernel
\begin{equation}
	K_{T2} \left( u \right) 
	= \left( \frac{1}{\sqrt{\pi A_2 \left( \tilde{\tau}_2 \right)} u} \right) \exp \left( - \frac{z^2}{4A_2 \left( \tilde{\tau}_2 \right) u} \right)
	\label{eq31}
\end{equation}
indicates that both molecular diffusion and thermo-acoustic energy transfers occur at the BL scale.
The expresion for  the expansion velocity which generates the heating acoutic field,  
\begin{equation}
	\begin{array}{rl}
		u_\infty \left( \theta \right) & = \mathop{\lim }\limits_{z \to \infty } u \left( z , \theta \right) = \bar{u} \left( \theta , x = 0 \right)
		\\
		& = \frac{2}{3} A_2 \left( \tilde{\tau}_2 \right) \tilde{\tau}_2^{-1.5} W \left( \theta \right) \, ,
	\end{array}
	\label{eq32}
\end{equation}
indicates that the PE fades away as $\tilde {\tau }_2^{1 / 2}$ when nearing the critical point which is however beyond any experimental capability in normally viscous supercritical fluids.

\end{document}